# SECURITY CHALLENGES IN MOBILE AD HOC NETWORKS: A SURVEY

Ali Dorri and Seyed Reza Kamel and Esmail kheyrkhah

Department of Computer Engineering, Mashhad branch, Islamic Azad University, Mashhad, Iran.

*ABSTRACT*

*MANET is a kind of Ad Hoc network with mobile, wireless nodes. Because of its special characteristics like dynamic topology, hop-by-hop communications and easy and quick setup, MANET faced lots of challenges allegorically routing, security and clustering. The security challenges arise due to MANET's self-configuration and self-maintenance capabilities. In this paper, we present an elaborate view of issues in MANET security. Based on MANET's special characteristics, we define three security parameters for MANET. In addition we divided MANET security into two different aspects and discussed each one in details. A comprehensive analysis in security aspects of MANET and defeating approaches is presented. In addition, defeating approaches against attacks have been evaluated in some important metrics. After analyses and evaluations, future scopes of work have been presented.*

## 1. INTRODUCTION

In these years, progresses of wireless technology and increasing popularity of wireless devices, made wireless networks so popular. Mobile Ad Hoc Network (MANET) is an infrastructure-independent network with wireless mobile nodes. MANET is a kind of Ad Hoc networks with special characteristics like open network boundary, dynamic topology, distributed network, fast and quick implementation and hop-by-hop communications. These characteristics of MANET made it popular, especially in military and disaster management applications. Due to special features, wide-spread of MANET faced lots of challenges. Peer to peer applications [1], integration with internet [2], security [3], maintaining network topology [4] and energy [5, 6] are some of the most important challenges in MANET. We presented an analysis and discussion in MANET challenges in our previous work [7].

In MANET all nodes are free to join and leave the network, also called open network boundary. All intermediate nodes between a source and destination take part in routing, also called hop-by-hop communications. As communication media is wireless, each node will receive packets in its wireless range, either it has been packets destination or not. Due to these characteristics, each node can easily gain access to other nodes packets or inject fault packets to the network. Therefore, securing MANET against malicious behaviours and nodes, became one of the most important challenge in MANET [8].

The aim of this paper is to provide a brief discussion and analysis on MANET security. Based on MANET characteristics we defined three important security parameters for MANET. In addition, two different aspects of MANET security are discussed in details. Furthermore, we presented an analysis and discussion in security attacks and defeating approaches. Moreover, the most effective defeating approaches for MANET and their limitations are introduced. Finally, some research directions are discussed. Rest of this paper is organized as follow: in section 2, three important security parameters in MANET are presented. Section 3 presents two important aspects of security with a discussion on their strategies. Three combinational challenges with security are presented in section 4. Section 5 presents our analyses and classifications on security of MANET and presents some research interest in security. Section 6 introduces open research issues and directions of researches in MANET security. Finally section 7 concludes the paper and introduces best ways to secure MANET and presents some future works.

## 2. Important Parameters In MANET Security

Because of MANET's special characteristics, there are some important metrics in MANET security that are important in all security approaches; we call them "Security Parameters". Being unaware of these parameters may cause a security approach useless in MANET. Figure 1 shows the relation between security parameters and security challenges. Each security approach must be aware of security parameters as shown in Figure 1. All mechanisms proposed for security aspects, must be aware of these parameters and don't disregard them, otherwise they may be useless in MANET. Security parameters in MANET are as follows:

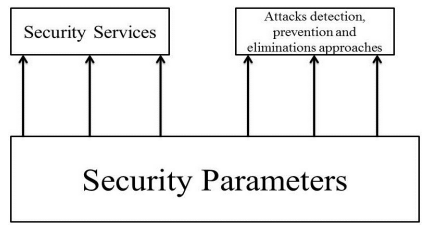

Figure 1. Relation between Security Parameters and Security aspects

**Network Overhead:** This parameter refers to number of control packets generated by security approaches. Due to shared wireless media, additional control packets may easily lead to congestion or collision in MANET. Packet lost is one the results of congestion and collision. Therefore, high packet overhead increases packet lost and the number of retransmitted packets. This will easily wastes nodes energy and networks resources.
**Processing Time:** Each security approach needs time to detect misbehaviours and eliminate malicious nodes. Due to MANET's dynamic topology it's strongly possible that routes between two different nodes break because of mobility. Therefore, security approaches must have as low as possible processing time in order to increase MANET flexibility and avoid rerouting process.
**Energy Consumption:** In MANET nodes have limited energy supply. Therefore, optimizing energy consumption is highly challengeable in MANET. High energy consumption reduces nodes and network's lifetime.
Each security protocol must be aware of these three important parameters. In some situations a trade-off between these parameters is provided in order to perform a satisfaction level in all of them. Security protocols that disregard these parameters aren't efficient as they waste network resources.

## 3. MANET Security Challenges

One of the earliest researches in security in MANET was presented in 2002 [9]. Some security challenges in MANET were inherited from ad hoc networks that were research interests since 1999 [10, 11]. Generally there are two important aspects in security: Security services and Attacks. Services refer to some protecting policies in order to make a secure network, while attacks use network vulnerabilities to defeat a security service. In the next two parts, a brief discussion on these security aspects is presented.

### 3.1. Security Services

The aim of a security service is to secure network before any attack happened and made it harder for a malicious node to breaks the security of the network. Due to special features of MANET, providing these services faced lots of challenges. For securing MANET a trade-off between these services must be provided, which means if one service guarantees without noticing other services, security system will fail. Providing a trade-off between these security services is depended on network application, but the problem is to provide services one by one in MANET and presenting a way to guarantee each service. We discuss five important security services and their challenges as follows:

**Availability:** According to this service, each authorized node must have access to all data and services in the network. Availability challenge arises due to MANET's dynamic topology and open boundary. Accessing time, which is the time needed for a node to access the network services or data is important, because time is one of the security parameters. By using lots of security and authentication levels, this service is disregarded as passing security levels needs time. Authors in [12] provided a new way to solve this problem by using a new trust based clustering approach. In the proposed approach which is called ABTMC (Availability Based Trust Model of Clusters), by using availability based trust model, hostile nodes are identified quickly and should be isolated from the network in a period of time, therefore availability of MANET will be guaranteed.

**Authentication:** The goal of this service is to provide trustable communications between two different nodes. When a node receives packets from a source, it must be sure about identity of the source node. One way to provide this service is using certifications, whoever in absence of central control unit, key distribution and key management are challengeable. In [13] the authors presented a new way based on trust model and clustering to public the certificate keys. In this case, the network is divided into some clusters and in this clusters public key distribution will be safe by mechanisms provided in the paper. Their simulation results show that, the presented approach is better than PGP. But it has some limitations like clustering. MANET dynamic topology and unpredictable nodes position, made clustering challengeable.

**Data confidentially:** According to this service, each node or application must have access to specified services that it has the permission to access. Most of services that are provided by data confidentially use encryption methods but in MANET as there is no central management, key distribution faced lots of challenges and in some cases impossible. Authors in [14] proposed a new scheme for reliable data delivery to enhance the data confidentially. The basic idea is to transform a secret message into multiple shares by secret sharing schemes and then deliver the shares via multiple independent paths to the destination. Therefore, even if a small number of nodes that are used to relay the message shares, been compromised, the secret message as a whole is not compromised. Using multipath delivering causes the variation of delay in packet delivery for different packets. It also leads to out-of-order packet delivery.

**Integrity:** According to integrity security service, just authorized nodes can create, edit or delete packets. As an example, Man-In-The-Middle attack is against this service. In this attack, the attacker captures all packets and then removes or modifies them. Authors in [15] presented a mechanism to modify the DSR routing protocol and gain to data integrity by securing the discovering phase of routing protocol.

**Non-Repudiation:** By using this service, neither source nor destination can repudiate their behaviour or data. In other words, if a node receives a packet from node 2, and sends a reply, node 2 cannot repudiate the packet that it has been sent. Authors in [16] presented a new approach that is based on grouping and limiting hops in broadcast packets. All group members have a private key to ensure that another node couldn't create packets with its properties. But creating groups in MANET is challengeable.

In previous part a brief discussion on security services and their challenges in MANET was provided. Detecting and eliminating malicious nodes, is another aspect of the MANET security. In the next section, important attacks in MANET and existing detection and/or elimination approaches to secure network against them is discussed.

### 3.2. Attacks

Due to special features like hop-by-hop communications, wireless media, open border and easy to setup, MANET became popular for malicious nodes. Some of the most important attacks in MANET are as follows:

**Black Hole Attack:** In this attack, malicious node injects fault routing information to the network and leads packets toward itself, then discards all of them [17-19]. In [20] we present a survey on black hole detection and elimination approaches. Also we presented a classification of defeating approach for this attack. Authors in [21] presented a new approach based on confirming the best path using second path. In this approach, whenever a source node receives RREP packets, it send a confirmation packet through the second best path to the destination and ask the destination whether it has a route to the RREP generator or to the Next_Hop_Node of RREP generator or not. If the destination has no route to this nodes, both RREP generator and it's Next_Hop_Node will mark as malicious nodes. Using this approach source node can detect cooperative malicious nodes. Whoever in the case of more than two cooperative malicious node, this approach can't detect all malicious nodes.

**Worm Hole Attack:** In worm hole attack, malicious node records packets at one location of the network and tunnels them to another location [22]. Fault routing information could disrupt routes in network [23]. Authors in [24] presented a way to secure MANET against this attack by using encryption and node location information. But as mentioned before, key distribution is a challenge in MANET.

**Byzantine attack:** In this attack, malicious node injects fault routing information to the network, in order to locate packets into a loop [25, 26]. One way to protect network against this attack is using authentication. Authors in [27] presented a mechanism to defeat against this attack using RSA authentication.

**Snooping attack:** The goal of this attack is accessing to other nodes packets without permission [28]. As in MANET packets transmitted hop by hop, any malicious node can capture others packets.

**Routing attack:** In this attack, malicious node tries to modify or delete node's routing tables [17, 18, 29]. Using this attack, malicious node destroys routing information table in ordinal nodes. Therefore, packet overhead and processing time will increase.

**Resource consumption attack:** In this attack, malicious node uses some ways to waste nodes or network resources [30, 31]. For instance, malicious node leads packets to a loop that consists of ordinal nodes. As a result, node's energy consumed for transmitting fault packets. In addition, congestion and packet lost probability will increase.

**Session hijacking:** Session hijacking is a critical error and gives an opportunity to the malicious node to behave as a legitimate system [32, 33]. Using this attack, malicious node reacts instead of true node in communications. Cryptography is one of the most efficient ways to defeat this attack.

**Denial of service:** In this attack, malicious node prevents other authorized nodes to access network data or services [34-38]. Using this attack, a specific node or service will be inaccessible and network resources like bandwidth will be wasted. In addition, packet delay and congestion increases.

**Jamming attack:** Jamming attack is a kind of DOS attack [39]. The objective of a jammer is to interfere with legitimate wireless communications. A jammer can achieve this goal by either preventing a real traffic source from sending out a packet, or by preventing the reception of legitimate packets [40].

**Impersonation Attack:** Using this attack, attacker can pretend itself as another node and injects fault information to the network [41-43]. As MANET has open border and hop-by-hop communications, it's hardly vulnerable against this attack. In some cases even using authentication is useless.

**Modification Attack:** In this attack, malicious nodes sniff the network for a period of time. Then, explore wireless frequency and use it to modify packets [44, 45]. Man-in-the-middle is a kind of Modification attack.

**Fabrication Attack:** In fabrication attack, malicious node destroys routing table of nodes by injecting fault information [46-48]. Malicious node creates fault routing paths. As a result, nodes send their packets in fault routes. Therefore, network resources wasted, packet delivery rate decreased and packet lost will growth.

**Man-in-the-middle attack:** In this attack, malicious node puts itself between source and destination. Then, captures all packets and drops or modifies them [49-51]. Hop by hop communications are made MANET vulnerable against this attack. Authentication and cryptography are the most effective ways to defeat this attack.

**Gray Hole Attack:** This attack is similar to black hole. In black hole, malicious node drops all packets, while in this attack, malicious node drops packets with different probabilities [52-55]. As it relays some packets, detecting this attack is more complicated than black hole and some detection approaches like sniffing or watchdog will be useless in it.

**Traffic Analyse Attack:** The goal of this attack is sniffing network traffic to use them in another attack or in a specific time [44, 56]. Malicious node captures all packets to use them later. In this section we discussed security aspects in detail. Figure 2 presents a summarization of MANET's security aspects.

## 4. INCORPORATING SECURITY AND OTHER CHALLENGES

One way to provide security in MANET, besides decreasing network overhead, is to incorporate security approaches with other challenges. In this way, both challenges are solved by improving security parameters in total. We discuss these combinational approaches as follows:

**Secure routing protocols:** The aim of these approaches is to provide security in routing phase. When a node wants to create a path to a destination, it uses some mechanisms to find a secure path and detect malicious nodes in the selected path before sending packets or after sending a number of packets. Authors in [57] presented a secure routing protocol based on using IPSEC in MANET routing protocols. Authors in [58] presented a trust based security routing protocol to create secure path. In MANET, there is more than one path between two different nodes. Selecting best path based on both routing and security, will improve security parameters.

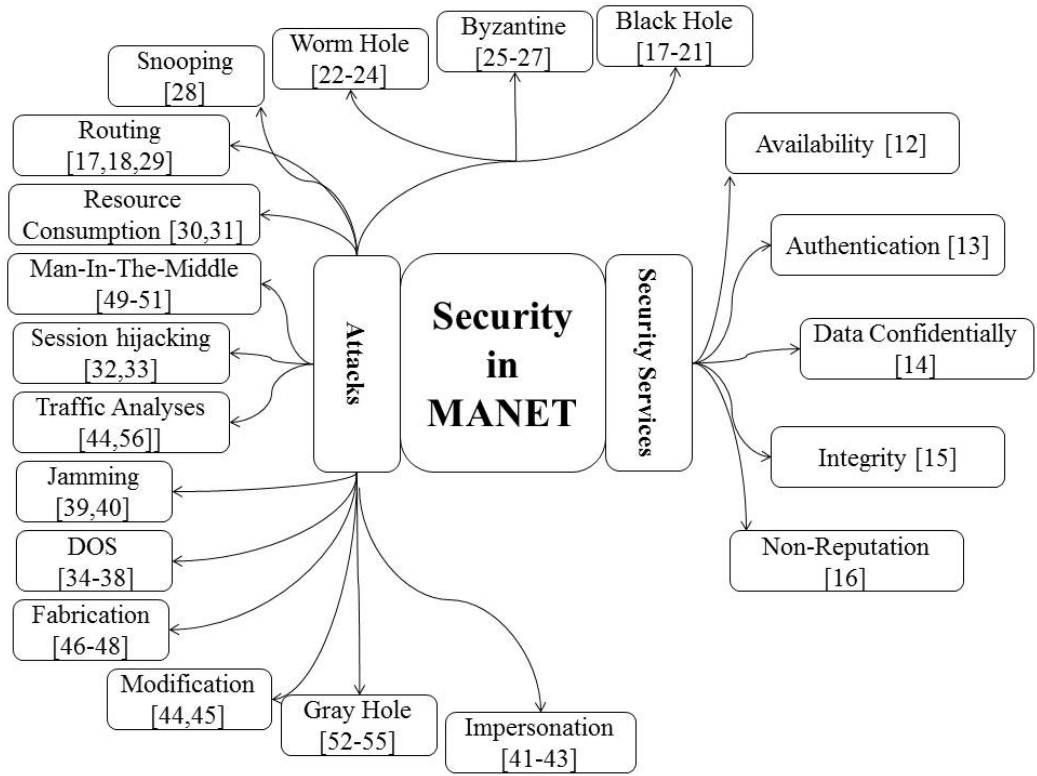

Figure 2. Security Aspects in MANET

**Security in QOS:** Using security mechanisms increase packet delivery time and processing time in each node. As a result, security has negative impacts on QOS. Therefore providing QOS beside security in MANET is highly challengeable. Authors in [59] presented a game theory to make a trade-off between security and QOS. Authors in [60] provided an approach that creates QOS aware multipath between source and destination with link information. By providing security in QOS, a level of security and QOS will be guaranteed with low time or network overhead.

**Cluster-based Security:** These approaches use clustering in order to provide more efficient situations for security protocols. Generally in these approaches, clusters are used for key distribution or as central management for controlling other nodes. Using clustering for security goals is important as it could solve problems in key distribution or key management. Authors in [61, 62] provides a key distribution mechanism by using clusters. Clustering can solve some of security challenges but cluster creation and maintenance is highly challengeable due to MANET's dynamic topology.

In incorporating approaches based on application, one challenge may be more important than the other one. For example, when selecting best path in routing is not as important as security, an approach can choose more secure paths without emphasizing on best routing path. As an example, in AODV routing protocol, the path with low sequence number is chosen as the best path. The reason is that, if there be a malicious node, it will send high sequence number. Using these benefits of incorporating approaches, both challenges will be solved with better security parameters.

## 5. ANALYSES AND DISCUSSION

Previous sections discussed incorporating security challenge with other challenges. This section presents an analytics and classification on previous issues. In order to analyse attacks and their behaviour, an analyses in each attack is presented in Table 1. For each attack five important parameters has been discussed. These parameters are as follows:

- Violated Service: Each attack breaks a security service. We presented the most important defeated service in this column.
- The Proposed Solutions: Some of the most effective approaches to detect and eliminate malicious nodes.
- MANET features which lead to this attack: Each malicious node uses a feature or features of MANET to break the security.
- Attack Type: Lots of researches classified attacks in two mainly class that are as follows: Active attack, Passive attack. In passive attacks, malicious node listens to transmissions without any active injection or effect on network [63]. While, in active attacks malicious node inject information.
- Attack Goal: The most important goal of each attack.

Referring to Table 1, it's understandable that lots of attacks are against availability security service. In availability aimed attacks, malicious nodes inject fault routing information or destroy nodes routing tables in order to defeat availability. Malicious nodes break availability service using passive or active attacks and in different types of attacks. For instance, in worm hole attack malicious node break availability by consuming resources, while in DOS attack, it drops all received packets in order to breach availability.

As another point in Table 1, it is clear that among defeating approaches, sniffing is one of the widely used approach in order to secure MANET. Generally in sniffing approaches, sniffer put itself in promiscuous mode and capture the network traffic. In this way, it can detect misbehaviour of malicious nodes. As another effective defeating approach, we can name encryption and route information. In encryption approaches, packet generator uses a key in order to encrypt packets to avoid malicious nodes from reading or modifying data packets. Whoever, it can discard encrypted packets. Therefore, encryption can't guarantee the availability service.

Routing information defeating approach uses route tables or additional control packets to detect malicious nodes. Refers to Table 1 this approach mostly used for active attacks. The reason is that in active attacks, malicious node injects packets or uses other nodes tables to break into the network.

Refer to Table 1, "Distributed network" is the most challengeable feature of MANET. Referring to this feature, in MANET all nodes cooperatively work with each other to configure and maintain network since there is no central control unit. Due to this feature, MANET faced lots of attacks since malicious nodes can pretend themselves as ordinal nodes and take part in network configurations and routing discoveries.

There are six different defeating mechanisms as presented in Table 1. By using defeating approaches all attacks detected and malicious nodes eliminated from the network. Therefore we present analyses in defeating approaches in Table 2.

Each proposed solution uses some of network resources in order to detect an attack. Table 2 presents' five important parameters for security approaches and discussed the effect of each solution in each parameter. In order to analyses the energy consumption of proposed solutions;

| Attack Name | Type | | Goal | | | | Violated Service | MANET features which lead to this attack | The proposed Solutions |
|---|---|---|---|---|---|---|---|---|---|
| | Active | Passive | Resource consumption | Accessing packets | Modification packets | Dropping packets | | | |
| Black Hole [17-21] | ✓ | | | | | ✓ | Availability | Distributed network | Routing information, Sniffing |
| Worm Hole [22-24] | ✓ | | ✓ | | | | Availability | Distributed network | Routing information, Encryption, Sniffing |
| Byzantine [25-27] | ✓ | | ✓ | | | | Availability | Distributed network | Encryption, Redundancy |
| Snooping [28] | | ✓ | | ✓ | | | Data Confidentiality, integrity | Non-centralized | Routing information |
| Routing [17,18],[29] | ✓ | | | ✓ | | | Availability | Hop-by-hop communications | Routing information, Authentication |
| Resource consumption [30,31] | ✓ | | ✓ | | | | ---------- | Non-centralized | Encryption, Sniffing |
| Session hijacking [32-34] | | ✓ | | ✓ | | | Data Confidentiality | Non-centralized Distributed network | Encryption, Authentication |
| Denial of service [35-38] | ✓ | | | | | ✓ | Availability | Non-centralized | Sniffing, Routing information |
| Jamming [39,40] | ✓ | | | | | ✓ | Availability | Wireless media | Sniffing, Dynamic frequency |
| Impersonation [41-43] | | ✓ | ✓ | ✓ | | | Data Confidentiality, Non-repudiation | Open network boundary | Authentication |
| Modification [44,45] | | ✓ | | | ✓ | | integrity | Hop-by-hop communication | Encryption |
| Fabrication [46-48] | | ✓ | ✓ | | ✓ | | Availability | Distributed network | Encryption, Sniffing |
| Man-in-the-middle [49-52] | | ✓ | | ✓ | ✓ | ✓ | Data Confidentiality, integrity | Hop-by-hop communication | Encryption, Authentication |
| Gray Hole [53-55] | ✓ | | | | | ✓ | Availability | Distributed network | Routing information, Sniffing |
| Traffic Analyze [44],[56] | | ✓ | | ✓ | | | Data Confidentiality | Hop-by-hop communication | Encryption, Authentication |

Table 1: Analytics on MANET Attacks.

Table 2 presents energy consumption of each approach in compare with other approaches. The word 'low' in "Energy consumption" column means referred approach consumes lower energy than other approaches.

Other words used in this column are "Normal" and "High" which mean normal and higher energy consumption in compare with other approaches. "Accuracy" column refers to ability of defeating approach to detect single or cooperative malicious nodes. In the case of cooperative attacks, malicious nodes work with each other in order to cover their tracks.

Routing information approach generates controller packets and uses them in order to detect malicious nodes. In addition, in some cases nodes must keep additional routing table like DRI table [19]. As mentioned, in sniffing approach each node must put itself in promiscuous mode and capture all packets transmitted in its range. This feature of sniffing approach wastes nodes energy. Also it increases process and memory overhead. In the case of cooperative malicious nodes, sniffing is useless as malicious nodes may work with each other to proof them-selves as trustable nodes.

Redundancy and dynamic frequency approaches can't detect the malicious nodes. These approaches can only avoid network from an attack. In the case of misbehaviour, these approaches can detect attack, while they are unable to detect the malicious nodes or eliminate them from whole network.

In routing information defeating approach, control packets transmission increase processing time of the security approach. In encryption and authentication, key distribution is an important challenge, because of lack of central infrastructure or key distribution centre. Therefore, each malicious node can pretend itself as a trustable node and take part in key distribution. In redundancy defeating approach, destination must buffer packets in order to get packets in sequence or to compare them with each other. In addition, it increases traffic overhead by sending duplicated packets. That cause increasing in congestion, packet lost and energy consumption.
In addition of five important parameters, Table 2 presents some limitations on each defeating approaches which are as follows:

- **Processing time:** This limitation refers to time needed for each approach to find malicious nodes and secure network against all malicious nodes. High processing time decreases MANET flexibility. The reason is that, during processing time of security approach routes must be constant and any change in route needs rerouting path, that increases time consumed for security approach.

Table 2: Analysis at The Proposed Solutions

| The proposed Solutions | Energy Consumption | Process Overhead | Memory overhead | Packet Overhead | Accuracy | Limitations |
|---|---|---|---|---|---|---|
| Routing Information [19],[25-27],[33] | Low | | ✓ | ✓ | S,C | Processing Time |
| Sniffing [22],[38],[56] | High | ✓ | ✓ | | S | Cooperative nodes |
| Encryption [49],[51],[61,62] | Normal | ✓ | ✓ | ✓ | S,C | Absence of Centralized Control, Key distribution |
| Redundancy [39],[60] | High | | | ✓ | None | Packet overhead |
| Authentication [41,42] | Low | ✓ | | ✓ | S | Absence of Centralized Control, Key distribution |
| Dynamic Frequency [39,40] | High | | | ✓ | None | Frequency knowledge |

- **Cooperative nodes:** In some cases defeating approach is unable to detect cooperative malicious nodes or it can detect with very high processing time. The reason of increasing processing time is that, the approaches can detect malicious nodes just one by one in different executions.
- **Key distribution:** Due to lack of central management in MANET, key distribution is challengeable in this network. Without key distribution centre, malicious nodes can get access to keys by capturing packets or by Man-In-The-Middle attack. One of the highly used approaches to overcome this challenge is using clusters in order to distribute keys in the network [64, 65].
- **Packet overhead:** This limitation refers to additional packets generated by the source node to detect and eliminate malicious nodes. Due to wireless media, high packet overhead in a defeating approach increases congestion, collision and packet lost probability in the network. In addition, it increases the processing time and energy consumption in the nodes.
- **Frequency knowledge:** Referring to this limitation, each node must know the frequency of transmissions. In MANET all nodes are free to join or leave the network, so being aware of all frequencies is highly challengeable.

Referring to our discussions and analyses it's understandable that routing information and encryption approaches are the most effective approaches for securing MANET. According to the application one of these two approaches is suggested. In cluster-based MANET using encryption is more effective and in MANET without any cluster using routing information is suggested.

## 6. FUTURE DIRECTIONS OF RESEARCHES

Until now we briefly discussed the security challenges in MANET and present some analytics in them. In this section we present open research issues.

Routing information approaches are suitable in all types of MANET. In this approach, reducing packet overhead and processing time, beside increasing accuracy is an important challenge. By increasing accuracy, it can detect cooperative malicious nodes. With decreasing processing time of this approach MANETs flexibility will increase.

Sniffing approach is useful in the case of single attacks, as it is unable to detect cooperative nodes. Whoever, it waste nodes energy and it is not suitable in MANET with high speed nodes.

Finding a more effective way to calculate the threshold and present effective detection mechanism forasmuch as decreasing time and packet overhead is the open border of research in sniffing approaches. Beside it, detecting cooperative malicious nodes is challengeable. In order to solve this challenge comparing sniffing with other defeating approaches is recommended.

MANET is self-organized, self-configurable network without any centralized control. Therefore, encryption and authentication are challengeable. Key distribution and control unit are the most important challenges. One way to over through these challenges is using clustering; therefore, the Cluster Head can act as the key distributer. Because of MANETs dynamic topology, creating and maintain clusters is highly challengeable. Using fuzzy logic [66] or swarm based [67] is highly recommended for this challenge. As another research interest, decreasing processing time and processing overhead of encryption approach can be mentioned.

Redundancy approaches, generate lots of duplicated packets and waste nodes resources. Also it increases congestion and packet lost. Effectively choosing number of duplicated paths, based on risk level, is highly challengeable. Also combining this approach with some other approaches in order to detect malicious nodes is another challengeable issue.
Dynamic frequency is effective in multi-type MANETs. By using this approach in multi-type MANET, each node secures its packets by sending in different frequencies. In addition, breaking one frequency has no effect on others. This is a challenge in this approach.

## 7. CONCLUSION

Mobile Ad Hoc Network (MANET) is a kind of Ad hoc network with mobile, wireless nodes. Due to its special characteristics like open network boundary, dynamic topology and hop-by-hop communications MANET faced with a variety of challenges. Since all nodes participate in communications and nodes are free to join and leave the network, security became the most important challenge in MANET.

In this paper, a comprehensive review in MANET security challenges is presented. Based on MANET characteristics and security requirements, three important security parameters are introduced. In addition, security divided into two different aspects and each one is briefly discussed. Furthermore, defeating approaches and different attacks in MANET are evaluated and analysed and future direction of work in each filed is introduced. Referring to our analyses and discussions, routing information and encryption defeating approaches are the most effective approaches for MANET security. Based on application one of these approaches can be used.

International Journal of Computer Science & Engineering Survey (IJCSES) Vol.6, No.1, February 2015

**Authors**

**Ali Dorri** received his B.S. degree in computer engineering from Bojnord University, Iran, in 2012, and now is student in M.S in software engineering in Mashhad branch, Islamic Azad University, Mashhad, Iran. His research interests cover Wireless Sensor Networks (WSN), Mobile Ad hoc Network (MANET) and specially Security challenges.

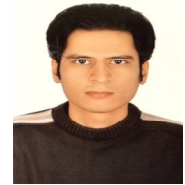

**Dr. Seyed Reza Kamel Tabbakh** is with the Department of Software Engineering, Faculty of Engineering, Islamic Azad University - Mashhad branch, Mashhad, Iran. He received his PhD in communication and network engineering from University Putra Malaysia (UPM) in 2011. He received his BSc and MSc in software engineering from Islamic Azad University, Mashhad branch and Islamic Azad University, South Tehran branch, Iran in 1999 and 2001 respectively. His research interests include IPv6 networks, routing and security. During his studies, he has published several papers in International journals and conferences. Email: rezakamel@ieee.org.

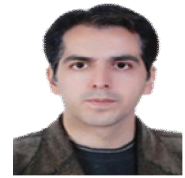

**Esmaeil Kheirkhah** received his Bachelor and Master in Computer Science and Mathematics from Islamic Azad University, Mashhad, Iran in 1992 and 1996 respectively. He also received his PhD in Computer Science from National University of Malaysia (UKM) in 2010. He is currently an assistant professor at the Islamic Azad University of Mashhad. His research interests include the Software Engineering, Requirements Engineering, End-User Computing, and semantic-enabled software engineering.

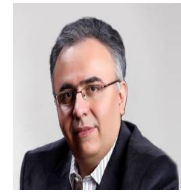